\documentclass[conference]{IEEEtran}
\IEEEoverridecommandlockouts
\usepackage{cite}
\usepackage{amsmath,amssymb,amsfonts}
\usepackage{algorithmic}
\usepackage{graphicx}
\usepackage{textcomp}
\usepackage{xcolor}

\usepackage{hyperref}
\usepackage{arydshln}  % For dashed line in tables
\usepackage{booktabs}
\usepackage{balance}
\def\BibTeX{{\rm B\kern-.05em{\sc i\kern-.025em b}\kern-.08em
    T\kern-.1667em\lower.7ex\hbox{E}\kern-.125emX}}
\begin{document}

\title{Toward Filament Segmentation\\Using Deep Neural Networks\\
% \thanks{Identify applicable funding agency here. If none, delete this.}
}

\author{
    \IEEEauthorblockN{
        Azim~Ahmadzadeh\IEEEauthorrefmark{1},
        Sushant~S.~Mahajan\IEEEauthorrefmark{2}, \\
        Dustin~J.~Kempton\IEEEauthorrefmark{1}, 
        Rafal~A.~Angryk\IEEEauthorrefmark{1} and
        Shihao~Ji\IEEEauthorrefmark{1}
    }
    \IEEEauthorblockA{
        \IEEEauthorrefmark{1}Department of Compute Science,
        Georgia State University,
        Atlanta, GA, USA\\
        \IEEEauthorrefmark{2}Department of Physics \& Astronomy,
        Georgia State University,
        Atlanta, GA, USA\\
        Email: \IEEEauthorrefmark{1}\{aahmadzadeh1,dkempton1,angryk,sji\}@cs.gsu.edu, \\
        \IEEEauthorrefmark{2}mahajan@astro.gsu.edu
    }
}

\maketitle

\begin{abstract}
We use a well-known deep neural network framework, called Mask R-CNN, for identification of solar filaments in full-disk H-$\alpha$ images from Big Bear Solar Observatory (BBSO). The image data, collected from BBSO's archive, are integrated with the spatiotemporal metadata of filaments retrieved from the Heliophysics Events Knowledgebase (HEK) system. This integrated data is then treated as the ground-truth in the training process of the model. The available spatial metadata are the output of a currently running filament-detection module developed and maintained by the Feature Finding Team; an international consortium selected by NASA. Despite the known challenges in the identification and characterization of filaments by the existing module, which in turn are inherited into any other module that intends to learn from such outputs, Mask R-CNN shows promising results. Trained and validated on two years worth of BBSO data, this model is then tested on the three following years. Our case-by-case and overall analyses show that Mask R-CNN can clearly compete with the existing module and in some cases even perform better. Several cases of false positives and false negatives, that are correctly segmented by this model are also shown. The overall advantages of using the proposed model are two-fold: First, deep neural networks' performance generally improves as more annotated data, or better annotations are provided. Second, such a model can be scaled up to detect other solar events, as well as a single multi-purpose module. The results presented in this study introduce a proof of concept in benefits of employing deep neural networks for detection of solar events, and in particular, filaments.

\end{abstract}


\section{Introduction}\label{sec:introduction}

    % Here we talk more generally about h-alpha images, and HEK system, why we use them and how we collect them. We should also talk about any pre/post processing phases we carry out, if at all.
    % Possibly, we can also talk about the challenged of working with this image source (BBSO).
    
    Our Sun has an extensive and complex magnetic field which has been studied for a century ever since George Ellery Hale \cite{1919ApJ....49..153H} discovered magnetic field in light coming from sunspots. This magnetic structure of the Sun is evident in many large-scale, as well as small-scale, features of the Sun because of the tendency of superheated ionized plasma to get trapped around strong magnetic field lines. One category of such large-scale features is \textit{solar filaments}.
    
    Filaments are accumulation of colder, denser plasma suspended in the solar corona along large-scale magnetic field lines in which the weight of the plasma is believed to be balanced by forces of magnetic origin. They are most clearly visible in Hydrogen and Helium (Lyman and Balmer) spectral lines. The availability of full disk H-$\alpha$ (a Balmer line) images of the Sun on a regular basis in which filaments appear as long dark threads against the solar disk facilitates their long-term studies. One such long-term collection of H-$\alpha$ images is available from the Big Bear Solar Observatory (1997-2019) \cite{denker1999synoptic} which is now a part of the Global High Resolution H-$\alpha$ network. These images are captured by filtering all light except the specific spectral line of H-$\alpha$ which is a deep-red visible spectral line with the wavelength of $656.28nm$. H-$\alpha$ images do not single out filaments as they also show \textit{sunspots}. It is, however, easy to visually differentiate between sunspots and filaments because sunspots have a round shape whereas filaments predominantly have an elongated, thread-like structure. Even though filaments constitute of plasma suspended in the solar corona, they are invariably found to be aligned with polarity inversion lines (PILs) over the solar surface (photosphere). PILs separate regions with opposite polarity large-scale magnetic flux on the photosphere. Filaments in the Corona have barbs (feet) extending down to the chromosphere and possibly connecting to the photosphere.
    
    % Why is detecting filaments important, problematic and challenging?
    
    \begin{figure}[t]
        \centering\includegraphics[width=\linewidth]{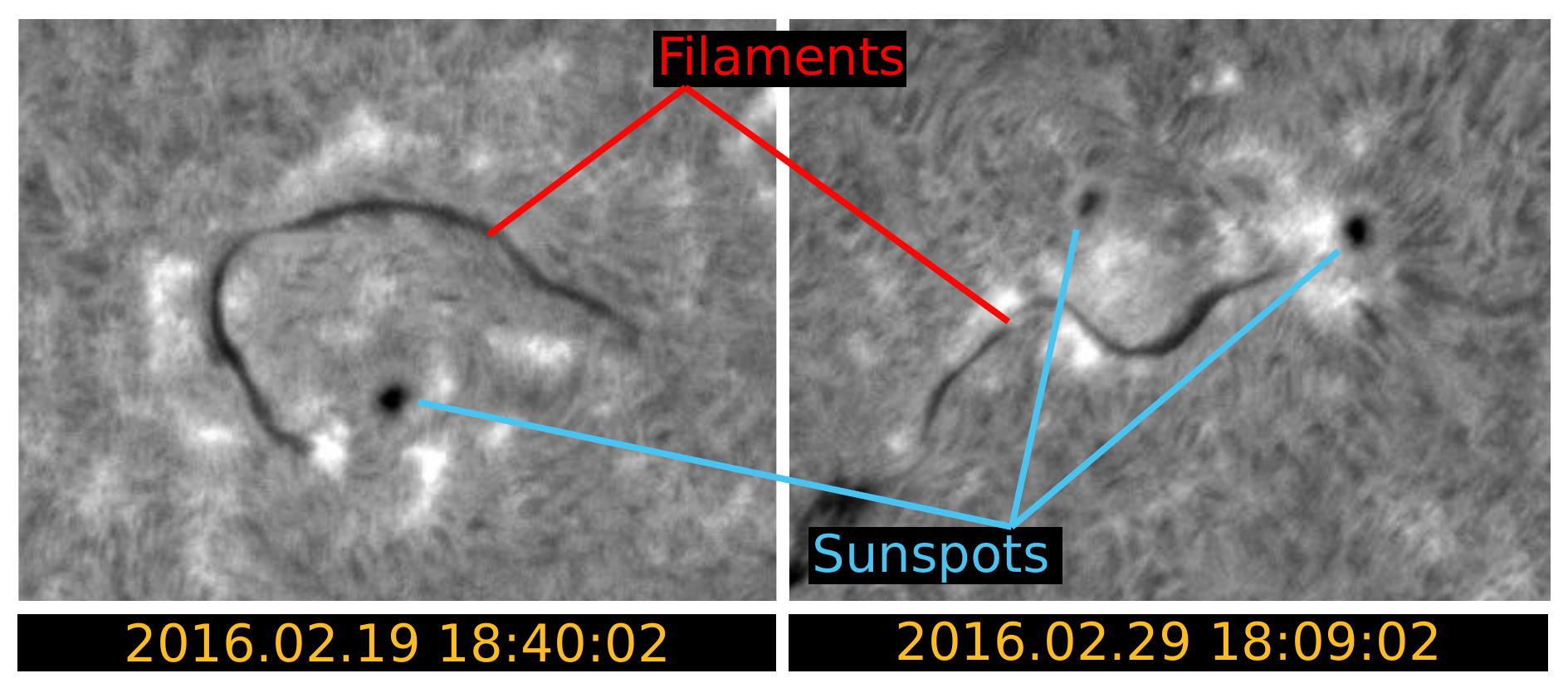}
        \caption{Several instances of filaments and sunspots, with very different shape structures, in two H-$\alpha$ images.}
        \label{fig:filamentSunspots}
    \end{figure}
        
    The prominent visual difference between filaments and sunspots is illustrated in Fig.~\ref{fig:filamentSunspots}. While this fundamental difference makes the classification of events fairly straightforward from the point of view of image processing algorithms, the background texture in H-$\alpha$ images is what serves as the most challenging part in segmentation problem. Presence of dark granularities in the background of H-$\alpha$ images makes it difficult, and sometimes impossible, even for human experts, to correctly differentiate between what is known as filaments' barbs and the background. This is specially important because one expectation from a reliable filament detection module is to characterize the shape and structure of the detected filaments and this can only be achieved in a high resolution segmentation of filaments that creates a pixel-level mask for each filament. In other words, determining an approximate vicinity of an event instance would not provide enough information for the scientific analysis of filaments. In this preliminary work, we exclude sunspot instances from our detection module, as we would like to start building a system from a single segmentation component, being filament detection, and upscale to more event instances, in the future.\par

    The automatic detection and characterization of solar filaments on a regular basis and for long term is performed by Bernasconi's \cite{bernasconi2005advanced} Advanced Automatic Filament Detection and Characterization (AAFDC) code which was a product of NASA's Feature Finding Team (FFT). This module finds one H-$\alpha$ image from BBSO every day and detects solar filaments in it using thresholding and image processing techniques based on visual features. A bounding box and a polygon which details the boundary of a filament detected by this module is reported to the Heliophysics Events Knowledgebase (HEK).
    
    In 2010, the Global High Resolution H-$\alpha$ network started capturing full disk H-$\alpha$ images of the Sun with a one minute cadence which is useful to study the dynamics of solar filaments including filament oscillations in order to study the processes which induct energy into filaments and to investigate triggering mechanisms which may be responsible for the sudden eruption of filaments. When filaments erupt, they may cause Coronal Mass Ejections (CMEs), which if directed towards the Earth can cause geomagnetic storms in polar regions on Earth, interfere with satellite communication, pose a threat to astronauts in space and induce currents in large-scale electrical grids on Earth.
    
    It is therefore important to develop a framework which can identify and characterize filaments in H-$\alpha$ images from observatories around the world at a cadence higher than a day. We in this paper present a step in this direction as we explore the use of deep neural networks (Mask R-CNN) for the segmentation (i.e. identification) of filaments.

% ----------------------------------------------------
%   II. DATA
% ----------------------------------------------------
\section{Data}\label{sec:data}
    
    \subsection{Data Sources}
        A large-scale analysis of solar events often requires two types of data: the observations, i.e., images, and the spatiotemporal metadata of the events of interest. Depending on the event-type of interest, there are a variety of instruments that provide images in different wavelength channels or with different filters that are more appropriate for some specific tasks. In addition to the needed image-types, the required resolution and the observation cadence can determine which telescope or instrument provides the most relevant data product.\par
    
        Full disk H-$\alpha$ images of the Sun are captured by multiple telescopes across the globe: the Big Bear Solar Observatory (BBSO) in California, the Kanzelh{\"o}he Solar Observatory (KSO)\footnote{\url{http://www.solobskh.ac.at}} in Austria, the Catania Astrophysical Observatory (CAO)\footnote{\url{http://woac.ct.astro.it/}} in Italy, Meudon\footnote{\url{http://bass2000.obspm.fr}} and Pic du Midi Observatories\footnote{\url{http://www.obs-mip.fr}} in France, the Huairou Solar Observing Station (HSOS) and the Yunnan Astronomical Observatory (YNAO)\footnote{\url{http://www.ynao.ac.cn/}} in China. In this preliminary work, we rely solely on the images provided by BBSO. The public archive of BBSO provides full-disk snapshots of the Sun in H-$\alpha$ filter, since 1997, on a daily basis. BBSO provides 2048$\times$2048-pixel images which are the highest in resolution compared to all other instruments producing a similar product. Since $6^{th}$ of July 2000, images in FITS format \cite{1981A&AS...44..363W} have also been added to the archive, in addition to the JPG format. This data format, in extension to the actual image in a 16-bit format, provides a vector of metadata such as the descriptive statistics derived from the pixel intensity distribution, the exact center and radius of the Sun corresponding to that image, the telescope configuration along with exposure time and wavelength filter used, quality of the image which for ground-based telescopes depends on the atmospheric seeing, and some further information which may be useful for accurate scientific use of the data.\par
            
        % Is this correct that FFT team is also responsible for detection of Filaments? Yes, it is.
        The spatiotemporal metadata of filaments (along with several other solar phenomena) detected by the Feature Finding Team (FFT) \cite{bernasconi2005advanced,martens2012computer}, are reported to the Heliophysics Events Knowledgebase (HEK) system \cite{hurlburt2010heliophysics} and can be accessed publicly through their API\footnote{\url{https://www.lmsal.com/hek/api.html}}. Among the numerous pieces of information accompanying each detected filament instance, in preparation for our segmentation task, we use the time of occurrences and the bounding boxes and polygons of the detected regions. A combination of the metadata and the actual images provides us with the information needed for training our filament-detection module.\par
        %Note that we are treating the output of the existing detection module as the ground truth\par
    
    \subsection{Data Acquisition}
        The full-disk H-$\alpha$ images were retrieved from the BBSO archive for the period of $2012$ through $2016$.
        For each day during this period, there exist multiple images available in both JPG and FITS formats. For this work the JPG format images would suffice, however, we still need the FITS files since their header information are needed for obtaining the best alignment of spatial objects with the events visible on the solar disk. Among these images, there are two variations: the raw images and those which have gone through flat field correction, dark subtraction and correction for limb darkening \cite{denker1999synoptic}. These corrections standardize most images taken on clear days. The day-to-day weather, however, is the dominant factor in the determination of the availability and the quality of images as even the presence of thin clouds in the sky can degrade the quality of observations well beyond the scope of corrections. Using these post-processed images reduces the computational load of our work and provides cleaner data for our neural networks.\par
        
        \begin{figure}[t]
            \centering\includegraphics[width=.8\linewidth]{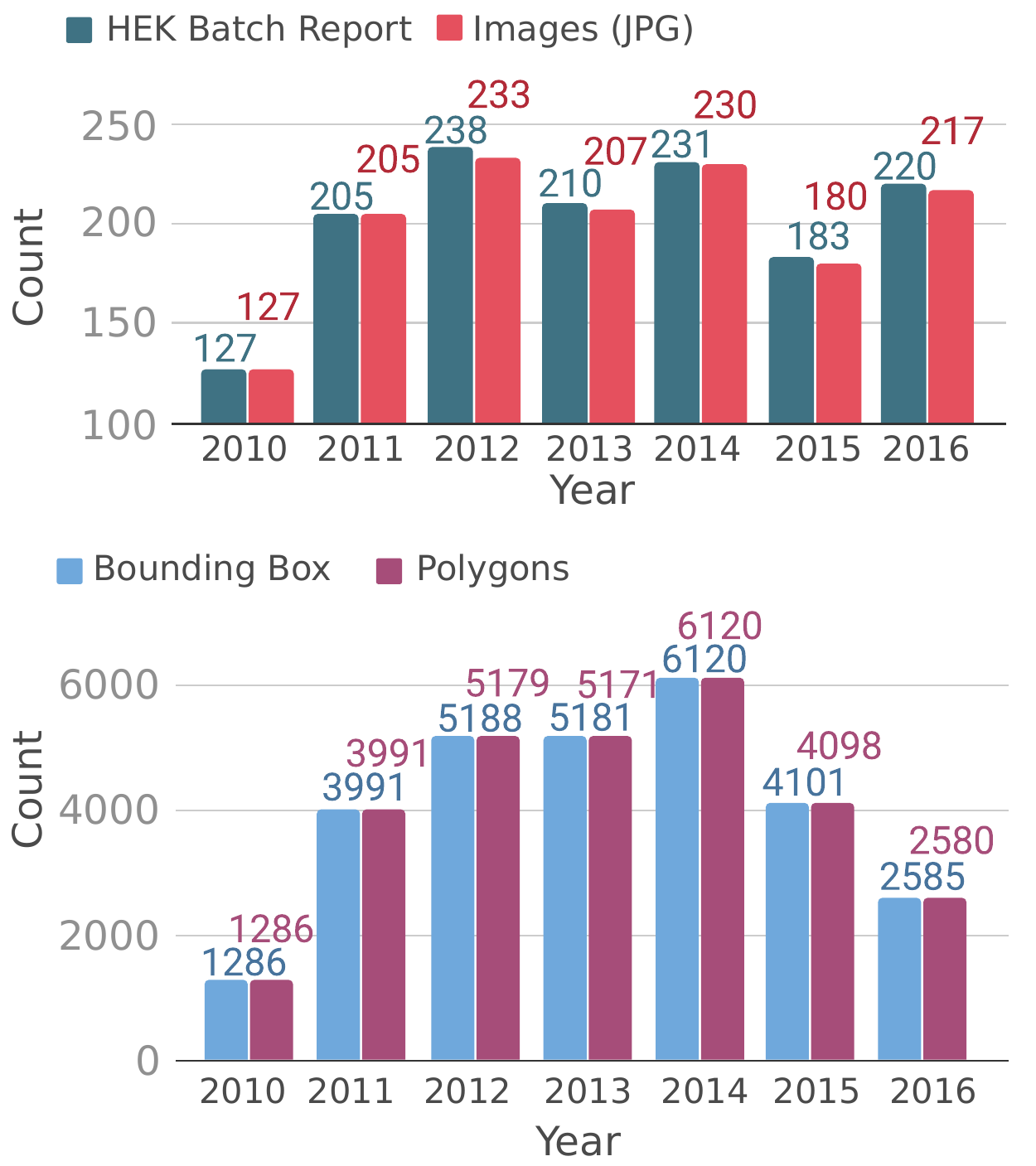}
            \caption{Comparison of the number of BBSO's H-$\alpha$ images and the HEK reports of filaments corresponding to those images (upper plot), and comparison of the number of filament bounding boxes and boundary polygons reported yearly to HEK (lower plot).}
            \label{fig:stats}
        \end{figure}
        
        Since the existing filament detection module (see Sec.~\ref{sec:stateoftheartModule} for more details about this module) attempts to detect all filaments present in each snapshot of the Sun at once, there is only one timestamp associated to all filaments present in one image. This timestamp does not represent any specific moment in the life-time of the filaments, such as the beginning of their formation, or when they are at their largest size during their evolution. And because this detection module uses the same data source (i.e., BBSO archive) for detection, the capture time of the images and the filaments' report time provided to HEK should match. To verify this, we first retrieve all filaments that are reported to HEK for a period of one year, and then search for the image in the archive that is temporally closest to this timestamp. With a tolerance of 3 minutes, we confirm that, except in a few cases during each year, the reports are spatially and temporally in line with the BBSO images. This is shown in Fig.~\ref{fig:stats} (upper plot).\par
        
        Similarly, we conduct a brief investigation on the correspondence of the reported bounding box of filaments and their corresponding boundary polygon. We noticed that in some cases the detected filaments are missing boundary polygons. The reverse situation, however, was not observed as all filaments had a bounding box. The overall analysis of such comparison is also depicted in Fig.~\ref{fig:stats} (lower plot).\par
        
        How many filaments are there for which neither a bounding box nor a polygon is generated by the existing module? This is a question that only a perfect filament detection module could answer. There are other questions of this nature as well that unfortunately we cannot confidently answer through an automated analysis. For instance, for how many filaments more than one polygons are generated? Or how often two spatially close filaments are identified as one single filament (or a filament channel)? Nonetheless, in the final section of our work, we compare the imperfection of our segmentation with the existing one's.\par
        
    \subsection{Data Integration}\label{subsec:dataIntegration}
        The data integration process for filaments is the process of mapping BBSO's image data to the spatiotemporal meta data about filaments provided to HEK. In other words, all filaments reported to HEK at time $t$ must be mapped to the image with the capture time $t \pm \tau$ where $\tau$ is the allowed time difference. We consider $\tau=3$ minutes to be an acceptable temporal tolerance rate given that the solar rotation period at the equator is $\approx 24$ days, and a $3$-minute rotation of the Sun is a relatively negligible shift, i.e., $\approx 0.5$ pixel and not even easily visible.\par

        The above spatiotemporal mapping allows us to build a dataset that is made of three parts: the H-$\alpha$ images in the JPG format, the image-specific meta data retrieved from the FITS's header, and the filament-specific spatiotemporal data. We've already discussed the first component. The header information, i.e., the second component, consists of the coordinates of the center of the solar disk, the apparent solar radius in pixels and the plate scale of each image\footnote{These values can be found in the header of BBSO's images, associated to the following keys: \texttt{CRPIX1}, \texttt{CRPIX2}, \texttt{IMAGE\_R0}, \texttt{CDELT1}, and \texttt{CDELT2}.}. Taking these values into account is crucial for the correct conversion of the spatial information down to the scale of the image. Filaments' bounding boxes and polygons are reported to HEK in arcseconds, considering the Sun's center as the origin of the space. To transform such information onto the pixel grid of the image with the origin being at the top-left corner of each image, the center of the Sun's disk and its radius are required at the image scale. Even though the Sun's radius does not physically change significantly, due to changes in the Sun-Earth distance, the apparent size of the Sun in these images and the physical size of one pixel in the image mapped onto the Sun changes throughout a year. If these changes are not taken into account, the alignment of the bounding boxes and filament boundary polygons will be significantly off on many occasions when they are overlaid on the actual filaments. The third component of our dataset is the filament-specific spatial data, i.e., the bounding boxes and polygons. For every image, a list of bounding boxes, each corresponding to one filament, is collected. These boxes are Minimum Bounding Boxes (MBR) enclosing the boundary polygons. They are defined with 5 points, starting and ending at the bottom-left corner of the box, ordered in the counter-clockwise fashion. A polygon is defined with a list of $n$ points, ordered similarly.\par
        
        In this study these spatial objects will be treated as the ground-truth data for our filament detection task since there is no sizable dataset of filaments available at this time, that is carefully annotated by the experts. Having said that, we are aware of the issues with the existing module, and we are not considering it as a perfect detection module. We are simply experimenting the extent to which a deep neural network can learn from the current detection methods with all evident strengths and weaknesses.\par
        
        \begin{figure}[t]
            \centering\includegraphics[width=\linewidth]{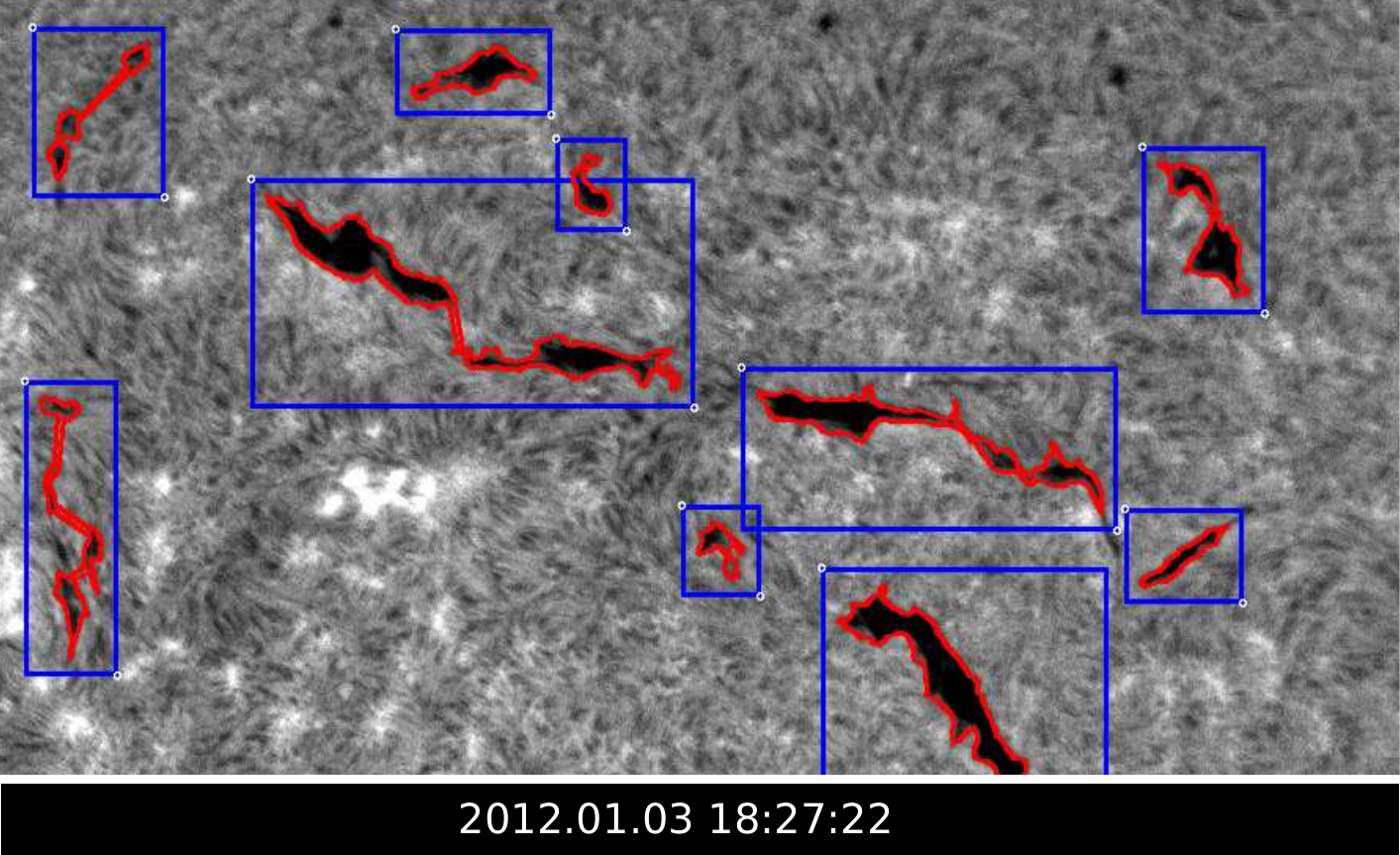}
            \caption{Visual verification of alignment of filaments as they appear in BBSO's H-$\alpha$ images, with their spatial bounding box (blue) and boundary polygon (red) information reported to HEK.}
            \label{fig:alignmentCheck}
        \end{figure}
    
    \subsection{Alignment Verification}\label{subsec:alignmentVerification}
        After the integration of all these three components is done, it is important to visually verify the alignment of the spatiotemporal data with the filaments visible in H-$\alpha$ images. This is mainly because simple mistakes in the integration process may result in arbitrary localization of filaments, and this could render the dataset entirely useless. The primary causes for these mistakes are as follows:
        \begin{itemize}
            \item Temporal mismatch of the reports: this could happen for a variety of reasons including the use of incorrect time zones when working with \textit{date-time}s from different databases with different assumptions in their design choices.
            \item Incorrect conversion from arcseconds to image pixel unit: mistakes such as incorrect assumption about where the origin is, or the order of $x$ and $y$ coordinates for each point, and also the order in which the polygon points form a shape, are some of the major causes of such potential misalignments.
        \end{itemize}
        
        A typical visual alignment verification analysis is shown in Fig.~\ref{fig:alignmentCheck}.
    
% ----------------------------------------------------
%   I. The State of the Art Module
% ----------------------------------------------------
\section{The Current State-of-the-art Module}\label{sec:stateoftheartModule}
    In 2005, a software for automation of detection and characterization of filaments was introduced by Bernasconi et al.~\cite{bernasconi2005advanced}, and became part of the solar event detection suite managed by the FFT. All pieces of data derived by this software from BBSO's H-$\alpha$ images are reported to the HEK system and therefore publicly accessible. The software is composed of four main components: image acquisition, image processing, filament detection and characterization, and filament tracking. In the first and second components, images are collected and then standardized by comparing the pixel intensity histogram of each image with a reference image. An image with good atmospheric seeing on a cloudless day is used as the reference. The third component is majorly responsible for creating the filament metadata. During this phase, a routine of classical image processing techniques with a sequence of thresholding filters is utilized to filter out the non-filament objects (both sunspots and some background noise) and create masks for objects which look like filaments. To get rid of some spurious pixels and small areas that are still among the objects flagged as potential filaments, eight morphological filtering operations are used to locate the small areas which are most certainly within the filaments' perimeter. These regions are then used as seeds to start a threshold-based clustering method for determining the filament masks. From this point on, a sequence of other techniques are used to build a profile for each filament. The profile describes each filament's unique structure, namely the main spine, the left and right barbs, and eventually the chilarity of the flux rope in which the filament is embedded. And finally, filaments individually and independently detected in sequential images are tracked in time while they travel across the visible solar disk.\par
    
    Bernasconi's algorithm for detection of filaments is diligent and in many cases, specially when the observations are very clear, has an overall good performance; the filaments are often spotted with an acceptable estimation and the chilarity corresponding to each filament that is calculated based on the detected characteristics of the barbs agrees with the experts' labels with a $72\%$ accuracy. Having said that, there is still a huge gap between the expected results and what the current module offers. While below we categorize the challenges present in the HEK's reports, we neither believe that these are necessarily the imperfections in Bernasconi's work (as in some cases, it is simply a design choice) nor we claim that our current work has overcome all these shortcomings. It is important, however, that we have a record of all potential defects as reported to HEK that we monitored. This work is motivated in this direction by these issues and we hope that such examples highlight the existing challenges.\par
    
    \begin{figure}[t]
        % Images used in this plot are listed below with the same order:
        % [GT_20140113193340.png][GT_20140223182451.png][GT_20140113193340.png]
        % [GT_20140211192301.png][GT_20140319175729.png][GT_20140322190141.png]
        % [GT_20140220190008.png][GT_20140225195954.png][GT_20140310180912.png]
        \centering\includegraphics[width=\linewidth]{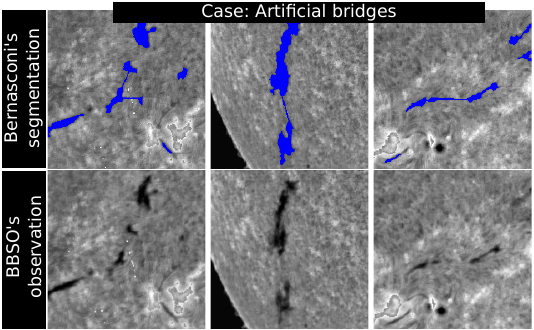}
        \centering\includegraphics[width=\linewidth]{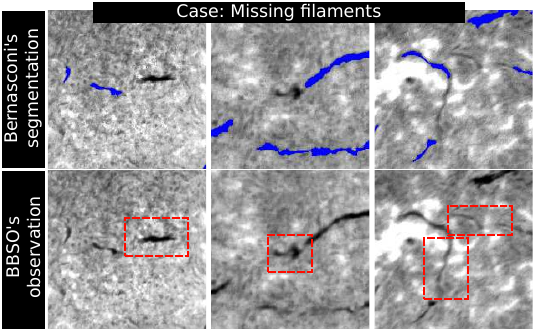}
        \centering\includegraphics[width=\linewidth]{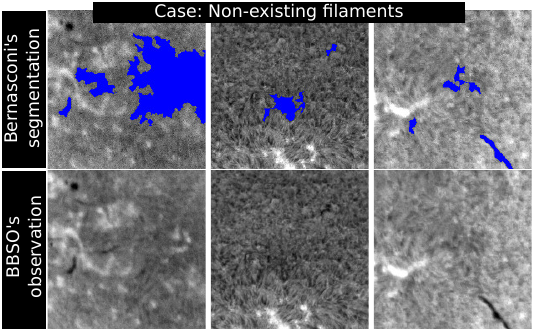}
        \caption{Three categories of typical misleading segmentations retrieved from HEK: artificial bridges, missing filaments, and non-existing filaments. The timestamps of these images, from top to bottom, left to right, is as follows: 20140113193340, 20140223182451, 20140113193340, 20140211192301, 20140319175729, 20140322190141, 20140220190008, 20140225195954, and $20140310180912$.}
        \label{fig:badSegmentations}
    \end{figure}
    
    A few examples of such cases are illustrated in Fig.~\ref{fig:badSegmentations}. The first case, depicts a behavior that is clearly a design choice: artificial bridges generated for filaments that are spatially close to each other, to report a filament channel, instead of multiple small filaments found in the same filament channel. Although, this decision is certainly of value for several queries, we believe that a filament detection module should remain independent of how different research objectives are defined. That is, it should only report what is observed, and leave the aggregation of filaments, if necessary, to the data-cleaning and pre-processing specific to each study. Having a data-driven model, as opposed to a task-driven model, is in particular important since different experts may have a slight disagreement on the pre-set proximity threshold that determines how nearby filaments could form a filament channel.\par
    
    The second case represents a few examples of where Bernasconi's model misses some filaments. Without a thorough investigation of every component of the software, it is not possible to confidently spot the issue, however, it seems that a combination of some extra constraints may have prevented such detections. There are many examples of such cases, present in almost every image, and it does not seem that reasons such as low contrast, corrupt images, or difficulties on detection closer to the limb of the Sun, could justify the majority of such examples. Having said that, the emphasized granularities in the background texture sometimes make it nearly impossible to spot small filaments, even manually. Another possible cause for such misses, might be rooted in the threshold-based filtering process used in the software, to get rid of the dark regions that are not filaments. As a result, some of the filaments might have been labeled as non-filaments by mistake, and thus removed from the remaining process.\par
    
    The third case is perhaps more interesting. Less often than the previous case, regions are annotated as filaments that are clearly not. The fact that in such cases, there are usually several other segmentations, eliminates the hypothesis of a general shift of segmentations due to time differences between the report and the timestamp of the image. It is more likely that, this is caused by incorrect choices of seeds in the process of threshold-based clustering. That is, small regions that are identified to be within filaments' regions, and then used as seeds, might have been chosen incorrectly because of a ``bad'' pre-defined threshold. The presence of thin clouds in the atmosphere at the time of the observation also interferes with the threshold based procedure for determining the seeds for filament masks.\par
    
    Of course, there could be a host of other reasons for such false negatives and false positives which are not revealed to us. Nonetheless, we use the same set of examples to show the differences and similarities between Mask R-CNN's segmentations and Bernasconi's, to avoid bias in our comparative analysis.\par
    
    Before we continue, let us clarify that all points made about Bernasconi's software and all comparisons presented here only concern the segmentation (i.e. identification) of filaments. Other components of the algorithm, such as profiling filaments based on their spines, spotting their left and right bearing barbs, and determining the chilarity of the filaments, are beyond the scope of this study and the presented model does not introduce any alternative for those functionalities, as we believe chilarity detection could be a different problem and requires a different approach.\par

\section{Neural Network Architecture}\label{sec:dnnArchitecture}
    Our filament detection problem stands as a specific application of the overarching object detection task which has been completely dominated by different deep neural network architectures, since 2009. AlexNet (2012) \cite{krizhevsky2012imagenet}, R-CNN (2014) \cite{girshick2014rich}, ResNet (2016) \cite{he2016deep}, and YOLO (2016) \cite{redmon2016you} are four of the most known models among many. In this work, we employ one of the improved versions of R-CNN, called Mask R-CNN (2017) \cite{he2017mask}, with ResNet-50-FPN backbone architecture. Mask R-CNN improves upon Faster R-CNN (2015) \cite{ren2015faster} by adding a branch for predicting segmentation masks on each RoI, which itself is a small Fully Convolutional Network (FCN). This results in a significant improvement on the main drawback of R-CNN, which is the inefficiency of the architecture that expects each image to be processed $\approx2000$ times. This is due to the use of the Selective Search algorithm \cite{uijlings2013selective} to obtain the region proposals and the fact that each proposed region should have been processed individually in the earlier models. Furthermore, they observed that the convolutional feature map can also be used for region proposal generation, which would make the entire system a single FCN.\par
    
    % TODO: The source code used needs to be cited

\section{Evaluation Metrics and Methodologies}\label{sec:evaluatioin}
    % http://cocodataset.org/#detection-eval
    % https://medium.com/@jonathan_hui/map-mean-average-precision-for-object-detection-45c121a31173
    % https://github.com/rafaelpadilla/Object-Detection-Metrics#different-competitions-different-metrics
    % https://www.import.io/post/history-of-deep-learning/
    Since the launch of ImageNet\footnote{\url{http://www.image-net.org/}} dataset in 2009 \cite{imagenet_cvpr09}, with more than $14$ million labeled images and more than $20,000$ categories, numerous detection models have been introduced. Following the ImageNet, competitions introduced their own challenges and for a consistent and fair comparison of different models, they each provided their own evaluation frameworks. After several years of exciting advancements in this area, the appropriate evaluation metrics for a general-purpose object detection model have become better understood and they generally converged to the metric set provided by Microsoft \cite{lin2014microsoft}. It was released as an API called cocoapi\footnote{https://github.com/cocodataset/cocoapi} along with a dataset and a series of competitions called Common Objects in Context (COCO)\footnote{\url{http://cocodataset.org/}}.\par
    
    Nonetheless, the objectives in a general-purpose object detection, and in particular segmentation, might be different than that in a specific domain such as the one we are pursuing in this work. In the former, a certain percentage of intersection between the ground-truth segmentation and the detected one could be considered as satisfactory. For instance, this is the case for spotting objects like Humans and Cars in images for the purpose of a real-time object tracking system. However, the goals might be set differently in other domains where geographical, medical, or astronomical images are the subject of study. In filament detection, as a relevant example, one of the segmentation applications would be to determine the chilarity of each filament based on the angle of their barbs against the main spine of the filaments \cite{pevtsov2003chirality}. This is simply not possible with a coarse segmentation. To this end, in addition to the well-known evaluation metrics provided by COCO API, we run our own analysis as well.\par
    
    \subsection{Average Precision and Average Recall}
        All metrics\footnote{See a list of all COCO's metrics and their definition at \url{http://cocodataset.org/\#detection-eval}} introduced by COCO API are derived from a measure called \textit{Intersection-over-Union} ($IoU$)\footnote{$IoU$ is more generally known as Jaccard Index, or Jaccard Similarity Coefficient.}, which is simply a normalized intersection of the ground-truth and the detected segmentation. More specifically, given $gt_i$ and $dt_i$ to be the \textit{ground-truth} and \textit{detected} segmentation, respectively, then $IoU_i = \tfrac{area(gt_i \cap dt_i)}{area(gt_i \cup dt_i)}$ quantifies the similarity or alignment of these two segmentations. In addition, a set of predefined thresholds over $IoU$ is required for determining true-positives and false-positives. This set is defined as $T = [0.5: 0.05: 0.95]$ which results in 10 different values for $IoU$ of each object. For instance, when the threshold is set to $0.8$, for each detection $i$ that $IoU_i \geq 0.8$, the detection counts as a true positive.\par
        
        \textit{Average Precision}, or $AP$, is an approximation of the area under the curve of \textit{precision} ($P = \tfrac{TP}{TP+FP}$) against \textit{recall} ($R = \tfrac{TP}{TP+FN}$). Since as the model progress in its classification of objects, recall always increases by occasional incorrect classifications, setting recall as the $x$-axis and monitoring the relative changes of precision could be summarized by the area under the curve. This value should be then averaged over all categories and depending on the chosen threshold $t \in T$, it can be denoted by $AP^{IoU=t}$. Following COCO's notation, $AP$, without any superscript, is also averaged over all 10 thresholds as well. In all COCO challenges, it is $AP$, averaged across all ten $IoU$ thresholds and all 80 categories, that determines the winner.\par
        % Do you want to talk about small-medium-large objects as well? If not, drop them from the table.
        
        \textit{Average Recall}, or $AR$, is the maximum achieved recall given a fixed number of detections per image, averaged over all categories and all IoUs. This is similar to what was proposed in \cite{DBLP:journals/corr/HosangBDS15}, except that in COCO it is averaged over all categories.\par
        
    \subsection{IoU Comparisons}
        For a more rigorous comparison of the detected and ground-truth segmentations, we should keep away from single quantities such as $AP$, and instead narrow down to a per-image analysis. To this end, we analyze $IoU$ of all pairs of $(gt_i$, $dt_j)$ in each image and look at the descriptive statistics. In this comparison, $i \in \{1, 2, \cdots, g\}$ and $j \in \{1, 2, \cdots, d\}$, where $g$ and $d$ are the total number of $gt$ and $dt$ segmentations, respectively, corresponding to that image. It is important to take into account only those pairs with non-zero intersections. The non-zero intersection constraint guarantees that only spatially relevant objects would be paired up. This is a helpful constraint based on the premise that the chances of $gt_r \cap dt_r = 0$, for the filament $r$, present in both detected and ground-truth sets, is very low. This is due to the fact that there is only one category (i.e., filaments) in our dataset and also owing to the flat nature of our images that filaments are not stacked over one another. In other words, if an annotated filament is detected, it will have some intersection with the ground-truth segmentation. In addition, this approach is insensitive to the missed objects. That is, neither a $dt$ with no matching $gt$, nor a $gt$ with no corresponding $dt$ segmentation would impact this metric. This is in particular important because the segmentations we considered as ground-truth are in fact another model's detection output and of course prone to minor or major mistakes. We refer to this methodology as \textit{pairwise comparison}, denoted by $IoU_{pairwise}$.\par
        
        \begin{figure*}[!htb]
            \centering\includegraphics[width=0.8\linewidth]{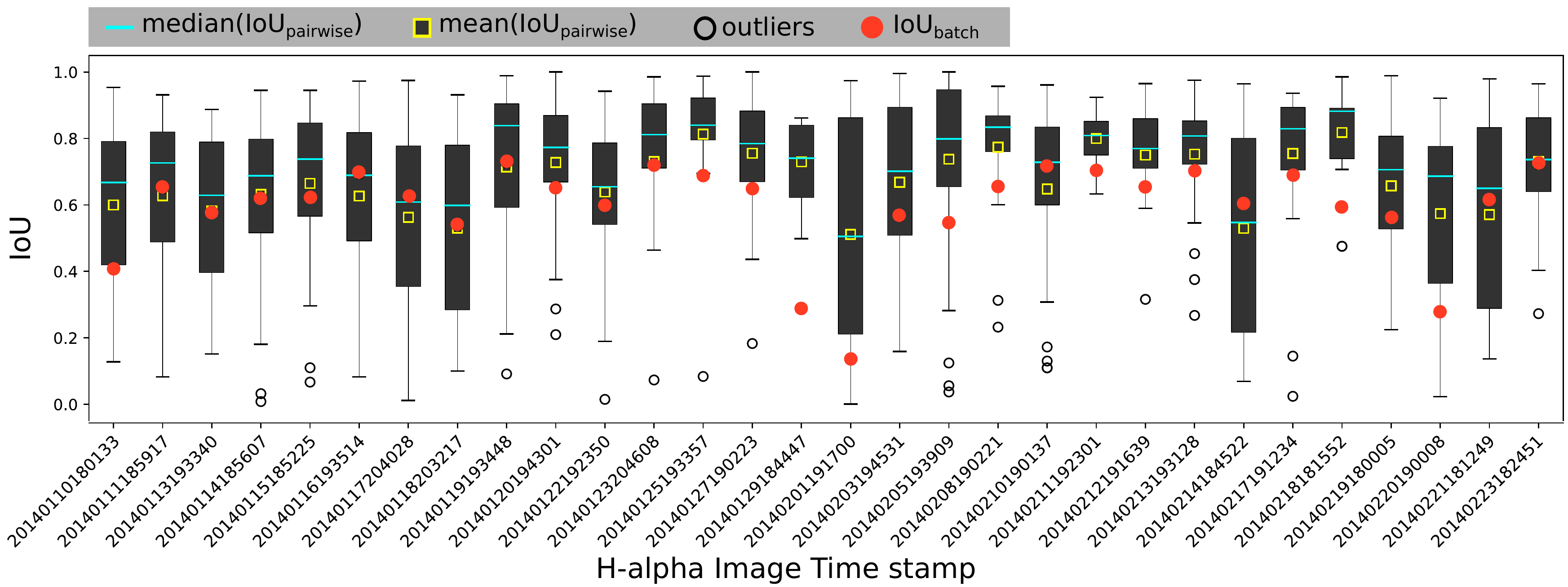}
            \caption{Box-plots of $IoU_{pairwise}$ for all filaments present on a collection of 30 images, as well as $IoU_{batch}$. The yellow squares show the mean value for \textit{pairwise comparison}s of all filaments in each image, that can be compared with the yellow crosses representing the \textit{batch comparison}s.}
            \label{fig:boxplots}
        \end{figure*}

        Having mentioned the advantages of this approach, it is important to understand its shortcomings as well. The main bias of this comparison is that it would be negatively impacted by the segmentations that may spatially agree with the ground-truth, but they differ in the number of pieces. In other words, a filament $i$ segmented as one piece, could be represented as the set $G = \{gt_i$\}, while this might be detected in $m$ smaller pieces, represented as $D = \{dt_{i1}, \cdots, dt_{im}\}$. Even if the area defined by $D$ and $G$ perfectly match (i.e., $IoU(D, G) = 1$), the above pair-wise comparison would result in multiple $IoU$s, one for each pair in $G \times D$. Each of those $IoU$s, however, are misleading quantities as they indicate a significant difference between smaller pieces in $D$ and $G$ which is a much larger area. This is in contrast to the fact that $D$ and $G$ perfectly match when compared collectively. This motivates us to aid our analyses with another approach that compensates for the above-mentioned bias. In this second approach, we group all $gt$ segmentations in one image, and form a single mask, denoted by $M_{gt}$. Similarly, using the $dt$ segmentations we create another mask, $M_{dt}$. Treating these masks as two objects, we can now compute $IoU(M_{gt}, M_{dt})$ which represents the quality of detection in each image. We refer to this comparison as \textit{batch comparison}, denoted by $IoU_{batch}$. While this approach perfectly avoids the above issue of the incorrect comparison of multi-piece segmentations, it would be directly affected by the missing segmentations in either the ground-truth or the detected sets. Looking at both of these measures together could give us a better insight into how similar our results are when compared to Bernasconi's detections.\par
        
\section{Results}\label{sec:results}
    
    \begin{table}
        \caption{Average Precision (AP) and Average Recall (AR) reports for filament detection and segmentation achieved by Mask R-CNN on BBSO H-$\alpha$ images.}
        \centering\footnotesize
        \begin{tabular}{l|l|l|l|l|l|l}
                \toprule
                \multicolumn{1}{c}{}  & \multicolumn{2}{c}{2014}
                    & \multicolumn{2}{c}{2015} & \multicolumn{2}{c}{2016}\\ \cdashline{1-7}[2pt/2pt]
                metric & bbox & segm  & bbox  & segm & bbox  & segm\\ \midrule
                $AP$            & 0.355 & 0.229 & 0.391 & 0.245 & 0.461 & 0.340\\ 
                $AP^{IoU = .5}$                      & 0.599 & 0.568 & 0.662 & 0.626 & 0.719 & 0.743\\ 
                $AP^{IoU = .75}$                     & 0.387 & 0.135 & 0.425 & 0.130 & 0.531 & 0.246\\ 
                $AR, max=1$     & 0.027 & 0.020 & 0.031 & 0.022 & 0.060 & 0.046\\ 
                $AR, max=10$    & 0.234 & 0.170 & 0.272 & 0.194 & 0.457 & 0.350\\ 
                $AR, max=100$   & 0.462 & 0.320 & 0.506 & 0.343 & 0.565 & 0.425\\ 
                \bottomrule
        \end{tabular}\label{tab:cocoEvaluation}
    \end{table}
        
    In this section, we analyze performance of Mask R-CNN on filament detection, in juxtaposition with Bernasconi's segmentations reported to HEK, using the metrics and methodologies discussed in Sec\.~\ref{sec:evaluatioin}. As the reader is looking at the results in this section, it is important to bear in mind a few key points in our experiments: (1) Mask R-CNN is employed as an off-the-shelf software without any hyper-parameter tuning necessary for approaching the best possible performance by the model on this specific dataset. We leave the tuning for our future work. However, (2) we do not use any pre-trained models. That is, all the weights are learned directly from the annotated filaments in BBSO images and no pre-trained model is used. Using pre-trained weights is a common practice, known as ``transfer learning'' \cite{pan2009survey}, that is in particular useful for general-purpose data such as Twitter text, Google images, etc. or the cases where there is some level of similarity, in terms of the patterns and structures, between the data used for learning and the data of interest. Most importantly, (3) the detection model employed in this study is intended to learn only from what is reported to HEK and no real ground-truth dataset, which is manually annotated by experts, is provided to it. In other words, we consider Bernasconi's segmentations on BBSO's H-$\alpha$ images as the ``ground-truth'' data to our training process. Although this limitation will impact the performance of the model, due to inheritance of at least some of the imperfections and weaknesses from the previous detection module, the extent of this impact should not be presumed without proper experiments. We investigate this impact in this section. Regardless, it is crucial to note that the model itself is completely independent from any detection module. That is, the utilized annotated data can be effortlessly replaced with any other and possibly less erroneously annotated data at any time, if provided.\par
    
    \begin{figure}[t]
        \centering\includegraphics[width=\linewidth]{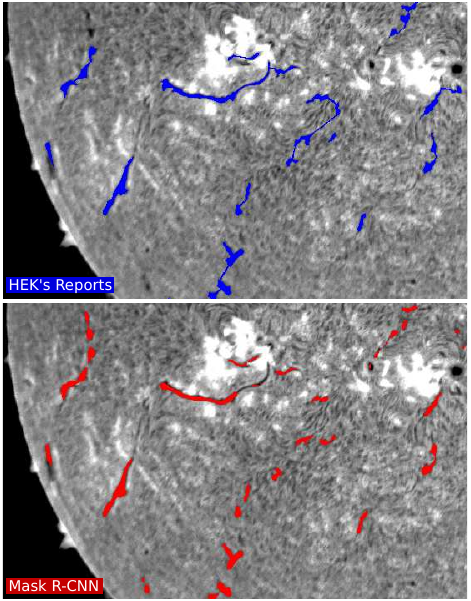}
        \caption{HEK's reports of filaments (top) and Mask R-CNN's segmentations (bottom), on a BBSO's image with timestamp 2014.02.14 18:45:22, corresponding to the box-plot with id `20140214184522' in Fig.~\ref{fig:boxplots}}.
        \label{fig:case1}
    \end{figure}
    
    Regarding the reported results in this section, we used one yer worth of data for training (2012), another year for validation (2013), and three other years of data (2014, 2015, and 2016) for testing. In all these three phases, we try detection with either bounding boxes or polygons reported to HEK. See Fig.~\ref{fig:stats} for the exact number of objects and images used in each phase.\par

    Table.~\ref{tab:cocoEvaluation} summarizes different $AP$ and $AR$ measures both for bounding box and segmentation detection. The results are reported for all BBSO's observations since 2014 through 2016. To better understand the numbers in the table, let us take the second row as an example and elaborate on it. The reported $AP$ indicates that the alignment of the detected segmentations with those annotated by Bernasconi's code, averaged over all images in 2016, with $IoU$ threshold fixed at $0.5$, is $0.743$. In other words, on average $\approx 74\%$ of all detected segmentations in this period have a relative overlap of $50\%$ or more with the ground-truth segmentations. To put this number in context, one could compare it with the best $AP$ achieved by a relatively similar architecture of Mask R-CNN trained and tested on COCO dataset, which is $58\%$ \cite{he2017mask}. Needless to say that our task, from the perspective of $AP$ and $AR$, is significantly simpler than the one put forward by COCO. For one, here we are dealing with one category, i.e., filaments, as opposed to the $80$ categories in COCO. However, the main challenge in our task, as we discussed in Sec.~\ref{sec:evaluatioin}, is the resolution of the segmentation and not distinction between different categories. This aspect of our problem is completely absent in tasks similar to COCO. This leads us to the other comparison methodologies as discussed in Sec\.~\ref{sec:evaluatioin}.\par
    
    \begin{figure}[t]
        \centering\includegraphics[width=\linewidth]{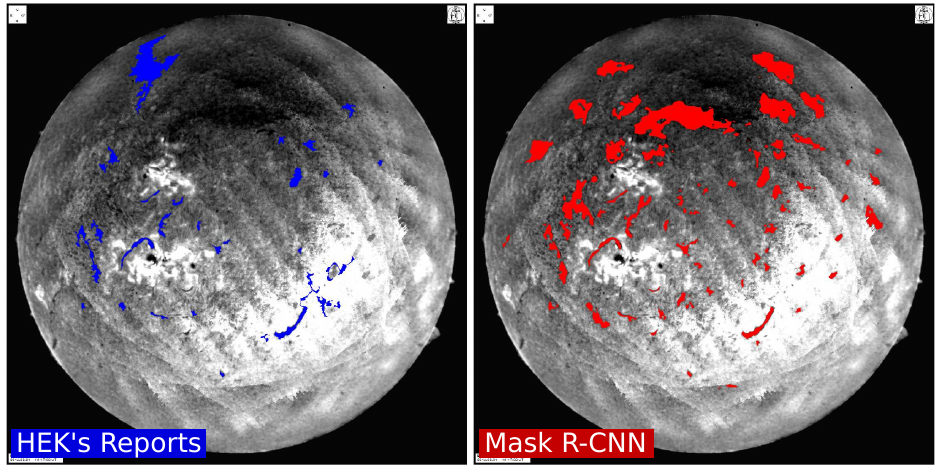}
        \caption{Impact of a highly defected observation on the segmentation task, with the timestamp 2014.02.01 19:17:00, corresponding to the box-plot with id `20140201191700' in Fig.~\ref{fig:boxplots}. This justifies the extremely low $IoU_{batch}$.}.
        \label{fig:case2}
    \end{figure}
    
    \begin{figure*}[t]
        \centering\includegraphics[width=\linewidth]{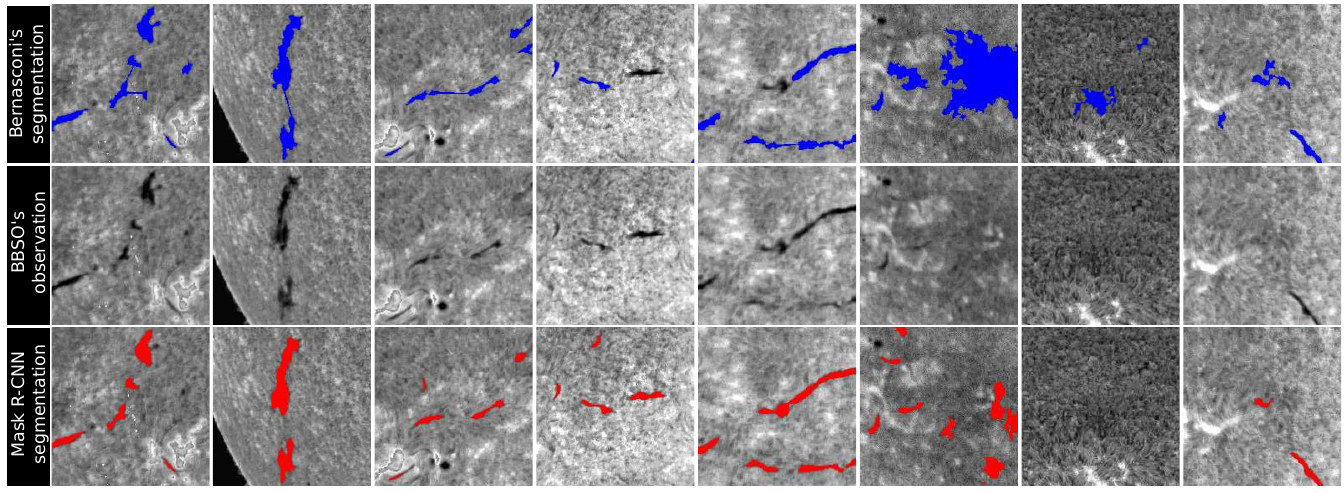}
        \caption{Comparison of Mask R-CNN's segmentation (top row, highlighted in blue) versus HEK's reports (bottom row, highlighter in red) on the same event instances used before in Fig.~\ref{fig:badSegmentations}.}
        \label{fig:finalComparison}
    \end{figure*}
    
    In Fig.~\ref{fig:boxplots}, we present the box-plots of $IoU_{pairwise}$ for 30 images, as well as $IoU_{batch}$, for each image. These images are selected randomly and the limited number of images allows a more visible visualization. While the degree of similarity to the Bernasconi's detection should not be considered as the objective, the plot shows that our model is overall in agreement with HEK's reports, with $IoU_{pairwise}$ averaging at $0.67$, slightly above the average of $IoU_{batch}$ at $0.59$. To obtain a better insight into these results, let us look into a few specific cases. One interesting case is the box-plot corresponding to the image id ending with `4522', that shows a relatively large interquartile range for $IoU_{pariwise}$. As both HEK's reports and our segmentations on this particular image are shown in Fig.~\ref{fig:case1}, there are several small dark regions that in HEK's reports are all connected with artificial bridges to form a filament channel, whereas in our segmentations, this is avoided. Although, this is simply a design choice, in our box-plot comparison this is reflected as high variance of $IoU$, but it should not be interpreted as an inaccurate segmentation. Another interesting case corresponds to the image id ending with `1700', where $IoU_{batch}$ is significantly low (i.e., less than 0.2). Tracking down the corresponding observation, shown in Fig.~\ref{fig:case2}, reveals the reason; the original BBSO's observation had produced a defected image based on which, any segmentation is spurious, hence very low alignment of segmentations. Our investigation shows that, Bernasconi's algorithm, due to a pre-set temporal requirement to observe the Sun at its highest elevation (lowest air mass) in order to maximize the quality of observations, used this corrupt image for segmentation while non-corrupt observations were available on that day. Other cases with very high $IoU$ and low variance, such as `3357' and `0221', are cases where the filaments are spotted against the less noisy background, and therefore the discrepancies are significantly less compared to some other cases. The outliers, shown as black circles in this plot, seem to be predominantly pointing out the comparison of a large, one-piece $gt$ segmentation with a relatively very small island in a multi-piece $dt$ segmentation.\par

\section{Conclusion and Future Work}\label{sec:conclusion}

    We have employed deep neural networks, in particular Mask R-CNN, for segmentation of filaments based on BBSO's full-disk H-$\alpha$ images and HEK's reports of filaments and their spatial information. We collected the data from BBSO's archive and integrated them with the spatiotemporal data retrieved form HEK to build our dataset in a way that it conforms to the COCO dataset format. We trained and validated our model on BBSO's observations during years 2012 and 2013, respectively, and tested it on three years worth of BBSO's archive, namely 2014, 2015, and 2016. We highlighted some typical and reoccurring segmentation characteristics of the existing detection module, and compared our findings with HEK's reports. Our case-by-case macroscopic study and the overall comparison of the two models show that (only in terms of segmentation of filaments) Mask R-CNN can clearly compete with the existing module and in some cases even performs better. This is an interesting outcome given that our model has only learned from what the existing module had detected and no actual ground-truth, i.e., data annotated by experts, were exposed to it.\par
    
    Our trained model, although still far from being robust and an operational software, encourages us to explore using deep neural networks for the detection of solar features. This argument is based on (1) As the model learned the important features, it has now become an independent tool that can function on any other observatories' data that has not been processed and annotated before. Given that the GONG full disk H-$\alpha$ network \cite{harvey1996global} provides images from observatories around the world (including BBSO). (2) The cadence of filament reports in HEK is daily as Bernasconi's code is designed to analyze one image per day. Due to our model being computationally inexpensive compared to Bernasconi's we can now provide filament reports with a cadence of one minute. Also, (3) such an automated system can be scaled up to cover all the solar events, instead of having one detection module specifically designed for each solar event. Moreover, (4) the performance of such a system is only bound to the amount of data provided to it. This is a well-known advantage of deep neural networks, as opposed to the classical image processing techniques or even shallow learning models whose performance is tied to the power of the features that are already selected.\par
    
    One of the avenues toward our future work, is to test this model on data from other observatories in the GONG full disk H-alpha network. We would like to see how different the performance of Mask R-CNN will be compared to segmentation on BBSO images, given that different instruments produce slightly different but comparable observations. In parallel, we plan to investigate on the possibility of increasing the resolution of segmentation taking into account the trade-off between adding more noise to the detected regions and the possibility of characterizing the filaments structure, i.e., the barbs and the spine, with a granularity comparable to the size of a pixel.\par

% \begin{table}[htbp]
% \caption{Table Type Styles}
% \begin{center}
% \begin{tabular}{|c|c|c|c|}
% \hline
% \textbf{Table}&\multicolumn{3}{|c|}{\textbf{Table Column Head}} \\
% \cline{2-4} 
% \textbf{Head} & \textbf{\textit{Table column subhead}}& \textbf{\textit{Subhead}}& \textbf{\textit{Subhead}} \\
% \hline
% copy& More table copy$^{\mathrm{a}}$& &  \\
% \hline
% \multicolumn{4}{l}{$^{\mathrm{a}}$Sample of a Table footnote.}
% \end{tabular}
% \label{tab1}
% \end{center}
% \end{table}

\section*{Acknowledgment}
    This work was supported in part by two NASA Grant Awards [No. NNH14ZDA001N], and one NSF Grant Awards [No. AC1443061 and AC1931555]. The AC1443061 award has been supported by funding from the Division of Advanced Cyber infrastructure within the Directorate for Computer and Information Science and Engineering, the Division of Astronomical Sciences within the Directorate for Mathematical and Physical Sciences, and the Division of Atmospheric and Geospace Sciences within the Directorate for Geosciences.\par

\balance
\bibliographystyle{IEEEtran}
\bibliography{main}

\end{document}